\documentclass[conference]{IEEEtran}
\IEEEoverridecommandlockouts

\usepackage{url}
\usepackage{xspace}
\usepackage{amsmath,amsfonts}
\usepackage{algorithmic}
\usepackage{graphicx}
\usepackage{textcomp}
\usepackage{xcolor}
\usepackage{tabulary}
\usepackage{caption}

\usepackage{tikz,pgf}

\usepackage{comment}
\usepackage{booktabs}
\usepackage{tikz}
\usepackage{pifont}
\usepackage{tablefootnote}
\usepackage{wrapfig}
\usepackage{enumitem}
\usepackage{paralist}

\usepackage{hyperref}
\usepackage{url}
\hypersetup{colorlinks,citecolor={blue}}

\usepackage{multirow}
\usepackage{subcaption}

\let\svtikzpicture\tikzpicture
\def\tikzpicture{\noindent\svtikzpicture}
\usepackage{listings}
\usepackage{adjustbox}

\newcommand{\xmark}{\ding{55}}
\newcommand{\fixeval}{\textsc{\textbf{FixEval}}\xspace}

\def\BibTeX{{\rm B\kern-.05em{\sc i\kern-.025em b}\kern-.08em
    T\kern-.1667em\lower.7ex\hbox{E}\kern-.125emX}}

\newcommand\scalemath[2]{\scalebox{#1}{\mbox{\ensuremath{\displaystyle #2}}}}

\definecolor{dark-gray}{gray}{0.85}
\definecolor{light-gray}{gray}{0.95}
\definecolor{mygreen}{rgb}{0,0.4,0}
\definecolor{mygray}{rgb}{0.5,0.5,0.5}
\definecolor{mymauve}{rgb}{0.58,0,0.82}
\definecolor{myred}{rgb}{0.82, 0.1, 0.26}

\usepackage{titlesec}
\titlespacing{\paragraph}{%
  0pt}{
  0.0\baselineskip}{
  1em}
  
\lstdefinestyle{CustomPy}{
    escapeinside={(*@}{@*)},
    belowcaptionskip=1\baselineskip,
    xleftmargin=1pt,
    xrightmargin=1pt,
    language=Python,
    numbersep=5pt,
    tabsize=4,
    showstringspaces=false,
    basicstyle=\footnotesize\sffamily, 
    keywordstyle=\color{mygreen},
    commentstyle=\color{purple},
    stringstyle=\color{red},
    identifierstyle=\color{black},
    numberstyle=\tiny\color{mygray},
    emph={int,char,double,float,unsigned,void,bool,boolean},
    emphstyle={\color{myred}},
    emph=[2]{and, in,},
    emphstyle=[2]{\color{violet}},
    emph=[3]{sortedCount, sorted_count},
    emphstyle=[3]{\color{blue}},
    numbers=left,
    stepnumber=1,
    breaklines=true,
    backgroundcolor=\color{white},
    literate={\ \ }{{\ }}1,
}

\lstdefinestyle{CustomJava}{
    belowcaptionskip=1\baselineskip,
    xleftmargin=1pt,
    xrightmargin=3pt,
    language=Java,
    numbersep=5pt,
    tabsize=4,
    showstringspaces=false,
    basicstyle=\footnotesize\sffamily, 
    keywordstyle=\color{mygreen},
    commentstyle=\color{purple},
    stringstyle=\color{red},
    identifierstyle=\color{black},
    numberstyle=\tiny\color{mygray},
    stringstyle=\color{mymauve},
    emph={int,char,double,float,unsigned,void,bool,boolean},
    emphstyle={\color{myred}},
    emph=[2]{and, in,},
    emphstyle=[2]{\color{violet}},
    emph=[3]{sortedCount, sorted_count},
    emphstyle=[3]{\color{blue}},
    numbers=left,
    stepnumber=1,
    breaklines=true,
    backgroundcolor=\color{white},
    literate={\ \ }{{\ }}1,
}
\makeatletter
\let\old@lstKV@SwitchCases\lstKV@SwitchCases
\def\lstKV@SwitchCases#1#2#3{}
\makeatother
\usepackage{lstlinebgrd}
\makeatletter
\let\lstKV@SwitchCases\old@lstKV@SwitchCases

\lst@Key{numbers}{none}{%
    \def\lst@PlaceNumber{\lst@linebgrd}%
    \lstKV@SwitchCases{#1}%
    {none:\\%
     left:\def\lst@PlaceNumber{\llap{\normalfont
                \lst@numberstyle{\thelstnumber}\kern\lst@numbersep}\lst@linebgrd}\\%
     right:\def\lst@PlaceNumber{\rlap{\normalfont
                \kern\linewidth \kern\lst@numbersep
                \lst@numberstyle{\thelstnumber}}\lst@linebgrd}%
    }{\PackageError{Listings}{Numbers #1 unknown}\@ehc}}
\makeatother

\newcount\myloopcounter

\newcommand{\repeatit}[2][10]{%
  \myloopcounter0
  \loop\ifnum\myloopcounter < #1 
  #2%
  \advance\myloopcounter by 1 %
  \repeat 
}

\begin{document}

\title{FixEval: Execution-based Evaluation of \\ Program Fixes for Programming Problems\\
 }

\author{
\IEEEauthorblockN{Md Mahim Anjum Haque}
\IEEEauthorblockA{
Virginia Tech \\
Blacksburg, VA, USA \\
mahim@vt.edu
}
\and
\IEEEauthorblockN{Wasi Uddin Ahmad}
\IEEEauthorblockA{
University of California, Los Angeles \\
Los Angeles, CA, USA \\
wasiahmad@ucla.edu
}
\and
\IEEEauthorblockN{Ismini Lourentzou}
\IEEEauthorblockA{
Virginia Tech \\
Blacksburg, VA, USA \\
ilourentzou@vt.edu
}
\and
\IEEEauthorblockN{Chris Brown}
\IEEEauthorblockA{
Virginia Tech \\
Blacksburg, VA, USA \\
dcbrown@vt.edu
}
}

\maketitle

\begin{abstract}
The complexity of modern software has led to a drastic increase in the time and cost associated with detecting and rectifying software bugs. In response, researchers have explored various methods to automatically generate fixes for buggy code. However, due to the large combinatorial space of possible fixes for any given bug, few tools and datasets are available to evaluate model-generated fixes effectively. To address this issue, we introduce \fixeval, a benchmark comprising of buggy code submissions to competitive programming problems and their corresponding fixes.
\fixeval offers an extensive collection of unit tests to evaluate the correctness of model-generated program fixes and assess further information regarding time, memory constraints, and acceptance based on a verdict. We consider two Transformer language models pretrained on programming languages as our baseline and compare them using match-based and execution-based evaluation metrics. Our experiments show that match-based metrics do not reflect model-generated program fixes accurately. At the same time, execution-based methods evaluate programs through all cases and scenarios designed explicitly for that solution.
Therefore, we believe \fixeval provides a step towards real-world automatic bug fixing and model-generated code evaluation. The dataset and models are open-sourced.\footnote{\url{https://github.com/FixEval/FixEval_official}}

\end{abstract}
\section{Introduction}
Repairing programs is one of the hardest and most expensive processes in software engineering. Finding and fixing errors, or debugging, takes up nearly $50\%$ of the total software development costs \cite{britton2013reversible} and 70-80\% of software engineers' time \cite{NIST}. Automation of program repair could enhance programmers' productivity \cite{seo2014programmers} and reduce software development costs \cite{goues2012systematic}.
Current research aims to provide solutions to automate program repair \cite{goues2019automated,gazzola2017automatic}. Automatic program repair is an active area of research\footnote{See \url{https://program-repair.org}} that can greatly relieve programmers from
the burden of manually fixing bugs in large codebases \cite{mesbah2019deepdelta, ding2020patching, dinella2020hoppity}. Researchers have recently started adapting language models to automate program repair tasks. Models such as PLBART~\cite{ahmad-etal-2021-unified} and CodeT5 \cite{wang2021codet5} have demonstrated promise in automating software engineering tasks, including automated program repair. 

While many approaches are being studied in the literature, more support is needed to better evaluate automated repair methods.
Prior work, such as TFix \cite{berabi2021tfix} and BIFI \cite{yasunaga2021break}, rely on n-gram metrics, e.g., BLEU or Edit Similarity. A limitation of n-gram metrics is that they penalize generated fixes if they differ from the reference, even if the fix is valid.
CodeBLEU \cite{ren2020codebleu} attempts to mitigate this by aggregating weighted n-gram, data flow, and syntax matches. While CodeBLEU is an improved strategy for evaluation, both n-gram metrics and CodeBLEU do not account for the large and complex space of program functionalities.
Therefore, a well-defined test suite is necessary \cite{Arcuri2008On, kim2013automatic,demarco2014automatic,ackling2011evolving} to evaluate a fix's correctness. 
A generated fix is considered functionally correct if it passes a set of unit tests. Overall, there is an increasing need for unit-test-driven benchmarks to assess program fixes generated by automatic program repair models.

In this work, we introduce \fixeval, a benchmark consisting of solutions to programming problems submitted by users in the AtCoder \cite{Atcoder_test_cases} and Aizu Online Judge~\cite{Aizu} platforms. Competitive programming requires programmers to attempt to solve tough problems within a specific time and memory limit. 
The process of competitive programming boils down to submitting code, receiving a verdict, making educated changes, and repeating these steps until an acceptable solution is reached. 
Thus, forming parallel examples of a buggy solution (not accepted) and a solution that passes a set of unit tests (accepted) could facilitate program repair evaluation. 
In \fixeval, we accompany buggy and correct program pairs with a suite of unit tests to evaluate the functional correctness of generated bug fixes. 
\fixeval contains solutions to $700$ programming challenges in both Python and Java and $25$ test cases on average per problem. We demonstrate the effectiveness of \fixeval through experimental evaluation and analysis of program repair techniques.

\noindent \textbf{Contributions:} The contributions of our work are summarized as follows: \textbf{(1)} We introduce \fixeval, a \textit{context-aware} program repair dataset that, along with unit tests, incorporates additional considerations---namely time and space complexity---to assess the functional correctness of repaired programs. \textbf{(2)} We benchmark the two most popular pretrained sequence-to-sequence language models PLBART and CodeT5 on \fixeval. \textbf{(3)} With thorough analysis, we demonstrate the necessity of unit-test-based program repair evaluation.

\begin{table*}[t!]
\centering
\resizebox{0.85\linewidth}{!}{
\renewcommand{\arraystretch}{1.1}
\begin{tabular}{l l l l l l}
\toprule
& \textbf{DeepFix}  & \textbf{Review4Repair} & \textbf{Bug2Fix} & \textbf{Github-Python} & \fixeval \\
\midrule
Language & C & Java & Java & Python & Java, Python\\
Dataset Test Size  & 6971 &  2961 & 5835, 6545 & 15k & 43k, 243k\\
Avg. \#Tokens & 203 & 320 + 37 & $ \leq $ 50, $ \leq $100 & \shortstack{ 10 - 128} & 331, 236 \\
Input Type & Program & Program + CR & Function & Program & Program \\
Error Type & CE Only & All  & All & CE Only &All\\
Test Cases & No & No & No & No & Yes \\
\bottomrule
\end{tabular}
}
\caption{
A comparison between \fixeval and other existing code repair datasets for machine learning. CR and CE indicate code review comments and compilation errors, respectively.}
\label{table:dataset_comparison}
\vspace{-2mm}
\end{table*}

\section{Related Work}

\subsection{Program Repair Benchmarks} 
A comparison between \fixeval and the recent program repair benchmarks is provided in Table \ref{table:dataset_comparison}. DeepFix~\cite{gupta2017deepfix} consists of approximately $7$K C programs written by students in an introductory programming course across $93$ programming tasks. 
However, DeepFix only covers compiler errors, does not provide test cases for evaluation, and fails to reflect real-world software applications. Review4Repair~\cite{huq2022review4repair} contains $55,060$ and $2,961$ examples of Java patches for training and evaluation, respectively. This work aims to repair code patches with the help of code reviews.
Bug2Fix~\cite{tufano2019learning} is a popular corpus used in CodeXGLUE~\cite{lu2021codexglue} that contains examples of the buggy and fixed Java code. However, the examples are at the function level, so cross-function dependencies are not modeled.
Bug2Fix also lacks unit tests to check for functional correctness. The GitHub-Python dataset~\cite{yasunaga2021break} is a collection of $38$K buggy and $3$M bug-free unparalleled code snippets from GitHub open-source Python projects. The $128$ token limit significantly reduces the overall problem complexity. However, the output code is defined as successful if it has no AST errors, which limits the focus only to compiler errors. 

Existing benchmarks, including test suites, have also been introduced to support automated program repair research. For instance, datasets such as IntroClass~\cite{LeGoues15tse} and Refactory~\cite{Refactory} consist of student assignments from introductory programming courses and provide unit tests. 
In contrast, QuixBugs~\cite{lin2017quixbugs} and Defects4J~\cite{just2014defects4j} provide real-world buggy programs with test suites. \fixeval is substantially larger than both datasets, consisting of more lines of code (QuixBugs: 1,034; Defects4J: 321,000; \textsc{FixEval}: 54M in Java and 61M in Python).
As a result, unlike existing benchmarks, \fixeval facilitates the training of machine learning models.
Moreover, QuixBugs only consists of programs with one-line defects, while Defects4J consists of Java code from five open-sourced programs. In comparison, \fixeval contains bugs that span multiple lines of code derived from 712K Java and 3.28M Python program submissions that vary in size and difficulty.  Although \fixeval was not collected from software repositories, we believe it is more representative for evaluating the capabilities of program repair models than the datasets above.


\begin{figure*}[t!]
\centering

\begin{tabular}{p{13.5cm}}
\toprule
\underline{Problem Statement}: 
A biscuit making machine produces $B$ biscuits at the following moments: $A$ seconds, $2A$ seconds, $3A$ seconds and each subsequent multiple of $A$ seconds after activation. Find the total number of biscuits produced within $T+0.5$ seconds after activation.\\
\underline{Constraints}: $1 \leq A,B,T \leq 20$, All input values are integers\\
\underline{Time Limit}: $2$ secs; \underline{Memory Limit}: $1024$MB; \underline{Problem Difficulty}: A\\
\bottomrule
\end{tabular}

\vspace{10pt}
\begin{tabular}{l}
\begin{adjustbox}{valign=t,minipage=0.47\textwidth}
\begin{center}
    \underline{Buggy Program in Java}
\end{center}
\begin{tabular}{l}

\lstset{escapechar=~,style=CustomJava}
\begin{lstlisting}[ 
    linebackgroundcolor={%
    \ifnum\value{lstnumber}=8
        \color{red!10}
    \fi
    }
]
import java.util.*;
public class Main {
    public static void main(String[] args){
        Scanner sc = new Scanner(System.in);
        int A = sc.nextInt();
        int B = sc.nextInt();
        int T = sc.nextInt();
        int S = T/A System.out.println(s*b);
    }
}
\end{lstlisting}
\end{tabular}
\end{adjustbox}
\hspace{10pt}
\begin{adjustbox}{valign=t,minipage=0.47\textwidth}
\begin{center}
    \underline{Fixed Program in Java}
\end{center} 
\begin{tabular}{l}
\lstset{escapechar=@,style=CustomJava}
\begin{lstlisting}[ 
    linebackgroundcolor={%
    \ifnum\value{lstnumber}=8
            \color{green!10}
    \fi
    \ifnum\value{lstnumber}=9
            \color{green!10}
    \fi
    }
]
import java.util.*;
public class Main {
    public static void main(String[] args){
        Scanner sc = new Scanner(System.in);
        int A = sc.nextInt();
        int B = sc.nextInt();
        int T = sc.nextInt();
        int S = T/A;
        System.out.println(s*b);
    }
}
\end{lstlisting}

\end{tabular}
\end{adjustbox}
\end{tabular}
\caption{Example submissions from the \fixeval dataset. Buggy and fixed statements are marked in red and green, respectively.}
\vspace{-2mm}
\label{tab:example_2}
\end{figure*}

\subsection{Program Repair Methods} 
Automated program repair aims to improve the debugging experience for developers by generating bug fixes automatically~\cite{goues2019automated}. Prior works in the literature model bug-fixing tasks in various ways. One of the most common approaches frames the task as machine translation from a buggy code to a fixed one.
Several researchers have shown language modeling is effective for automating coding tasks, such as program generation~\cite{ahmad-etal-2021-unified, wang2021codet5}, translation~\cite{ahmad-etal-2021-avatar}, and auto-completion~\cite{chen2021evaluating}. Nevertheless, there needs to be more research on applying language modeling in automated bug fixing and code repair.

\subsection{Evaluating Pretrained Language Models} 
Due to the recent success of large-scale language models in many domains~\cite{raffel2019exploring,brown2020language,shoeybi2019megatron}, new techniques have been introduced with pretraining objectives relevant to source code. Models such as BART~\cite{lewis2019bart}, GPT~\cite{chen2021evaluating}, and T5~\cite{raffel2019exploring} have been applied to software engineering tasks, demonstrating improvements in automating development tasks such as code generation, translation, bug detection, etc. For example, PLBART~\cite{ahmad-etal-2021-unified} uses denoising autoencoding to pretrain Transformer~\cite{vaswani2017attention} on programming language corpora. 
In contrast, TFix~\cite{berabi2021tfix} is a proposed method evaluating T5~\cite{raffel2019exploring} by leveraging commits from GitHub repositories to fix bugs.
We train a subset of these models on our dataset with various input configurations to evaluate their performance. 



\section{\fixeval Dataset}
\label{sec:dataset}

The \fixeval dataset contains parallel examples of the buggy and correct programs written in Java and Python languages. \fixeval examples are collected from CodeNet~\cite{puri2021project}. CodeNet dataset is a collection of programs written in over 50 languages by participants to solve competitive programming problems at the AtCoder and Aizu Online Judge platforms. \fixeval includes open-sourced unit tests for the AtCoder programming problems to facilitate evaluations to assess repaired programs' functional correctness.
Furthermore, \fixeval takes the time and memory limit set for each programming problem into consideration while executing repaired programs against the source code.


We argue that while competitive programming is not an exact reflection of real-world professional software development environments, by taking time and memory requirements into consideration, \fixeval extorts the practice of requiring software engineers to write efficient code in industrial settings~\cite{mens2012complexity}. Since \fixeval contains programs with different lengths and difficulties and is written by many programmers, \fixeval as a benchmark would foster future work on automatic program repairing.


\subsection{Dataset Construction} 
In the CodeNet dataset, every program is associated with a \texttt{user\_id} and \texttt{submission\_date}. We group the submitted programs by \texttt{user\_id} and then chronologically sort them. We refer to such chronological user submissions for a particular problem as a submission path.\footnote{From CodeNet, we extracted total $6.5$M submission paths.}
Each submission is associated with status, i.e., acceptance or error type.
We referred to the submission status as a \textit{verdict} for the submission.
If the submitted code passes all the hidden test cases, the verdict is \textit{Accepted (AC)}. Otherwise, submissions may receive a verdict from among 12 different options, the most frequent being: \textit{(i) Wrong Answer (WA)}, i.e., failed one or more test cases; \textit{(ii) Time Limit Exceeded (TLE)}, i.e., the program did not run within the intended time limit; \textit{(iii) Compilation Error (CE)}, i.e., the program did not compile; and \textit{(iv) Runtime Error (RE)}, i.e., program execution was not successful ~\cite{puri2021project}. 
For a user solving a particular problem, a sample submission path could be [WA, WA, TLE, AC] representing three failed submissions consisting of two incorrect codes and one inefficient implementation before arriving at the correct solution. 

We pair each unaccepted submission (i.e., a buggy code) from a user with the accepted submission (i.e., a bug-free code) to form an example (data point) in \fixeval. A sample example can be viewed in Figure \ref{tab:example_2}.
\fixeval consists of examples in Java and Python created from all $154$K users' submissions.
We de-duplicate the programs submitted for each problem using Jaccard similarity. We used the $javalang$\footnote{\url{https://github.com/c2nes/javalang}} tokenizer for Java and the $tokenizer$\footnote{\url{https://docs.python.org/3/library/tokenize.html}} standard library for Python.
As shown in Table \ref{tab:fixeval_stats}, we create stratified dataset splits based on problems to ensure a clear partition (80-10-10) in the train, test, and validation, with no overlapping problems or submissions across splits. We ensure all the test and validation data examples include test cases.


\subsection{Test Suite Collection}
We download all the publicly available test cases by AtCoder\footnote{\url{https://atcoder.jp/posts/21}} site. 
The test cases are manually created by domain experts (e.g., problem setters for the programming challenges) to cover all possible cases and ensure the functional correctness of the submitted programs. We also acknowledge the potential threat to validity as all the unit tests weren't manually validated. 
The test cases are organized into individual input and output files. The input files are directly used as input to a program. Correspondingly, the output of the executed program is compared with the output files.
To construct the \fixeval test suite, we match the problem names from AtCoder with names available as problem metadata in the CodeNet dataset. 
We manually remove some problems from \fixeval.
The problems are constraint satisfaction problems where the main goal is to satisfy conditions based on rules or designed constraints. Consequently, we also remove problems for which many combinatorial outputs are equivalently valid.
To tackle programs with 
numerical output, we assume a program is accepted if the difference between the program output and the test case output is below a certain precision threshold. All evaluations use the most frequent precision cutoff ($10^{-8}$).
On average, \fixeval validation and test set contain $24$ and $25$ test cases per problem. 

\begin{table*}[t!]
\centering
\resizebox{0.7\linewidth}{!}{
\begin{tabular}{l | cccc | cccc} 
\toprule
\multirow{2}{*}{Language} & \multicolumn{4}{c|}{Problem Count} & \multicolumn{4}{c}{Example Count} \\
\cmidrule(lr){2-5} 
\cmidrule(lr){6-9} 
& Train & Valid & Test & Total & Train & Valid & Test & Total \\
\midrule
Java &  2,160 & 279 & 279 & 2,718 & 156k & 44k & 43k & 245k \\
Python &  1,951 & 244 & 244 & 2,439 & 567k & 301k & 243k & 1,111k \\
\bottomrule
\end{tabular}}
\caption{
\fixeval dataset statistics.
}
\label{tab:fixeval_stats}
\vspace{-4mm}
\end{table*}



\begin{figure}[t!]
\centerline{\includegraphics[width=0.5\linewidth]{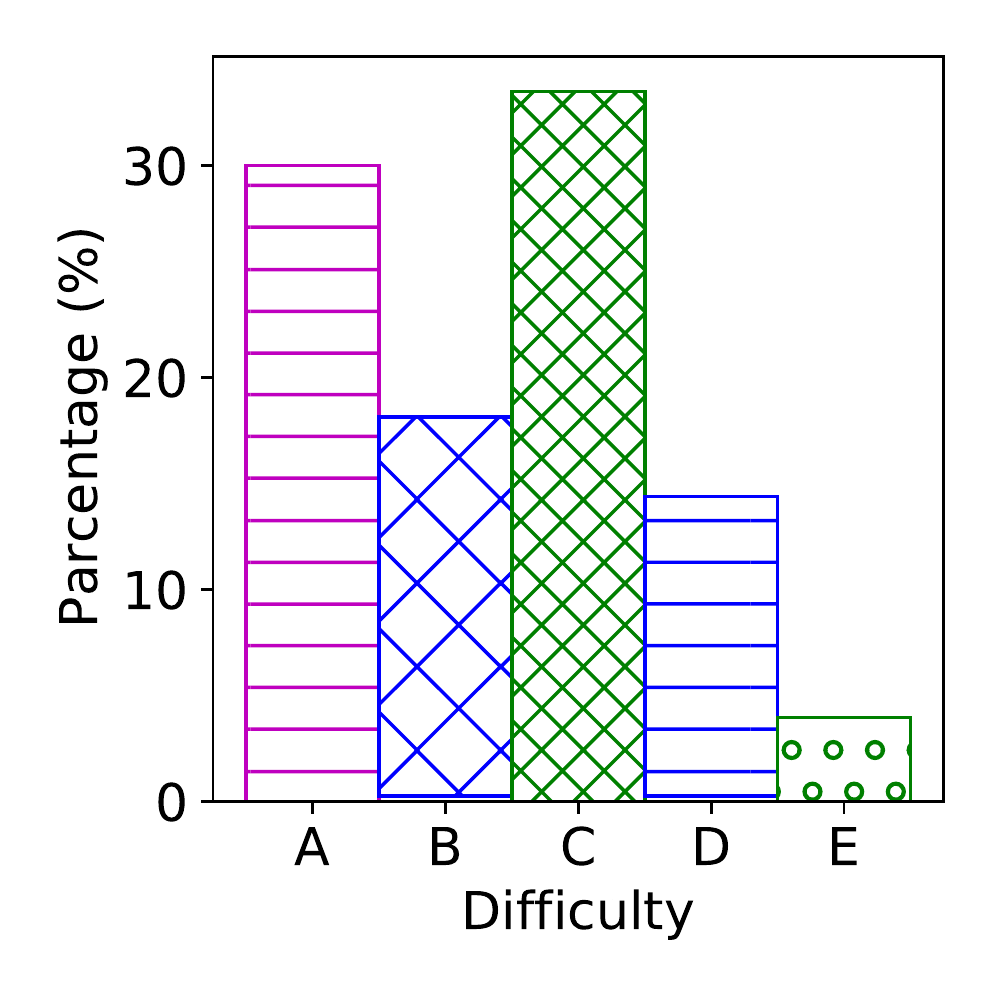}}
\vspace{-2mm}
\caption{Distribution of task difficulty (labels A to E indicate increasing difficulty) for the AtCoder problems belong to the test split in \fixeval.}
\label{fig:difficulty}
\vspace{-2mm}
\end{figure}

\subsection{Task Difficulty}
\label{subsec:difficulty}
The average length of the Java and Python programs in \fixeval are $331$ and $236$ tokens, respectively.
While the length of a program does not indicate the level of difficulty of repairing the program, we argue that since \fixeval programs' are larger than the existing program repair datasets (as shown in Table \ref{table:dataset_comparison}), \fixeval is a challenging benchmark for automatic program repair models.

Furthermore, we conjecture that the task labels assigned to AtCoder contest problems 
indicate problems' difficulty (labels A to E represent increasing difficulty).
In \fixeval, we retain the task labels and ensure the distribution of the labels across splits is even. The distribution of the task labels that we refer to as task difficulty for the test split is presented in Figure ~\ref{fig:difficulty}.



\section{Experiment Setup}

 \begin{table*}
\centering
\resizebox{0.75\linewidth}{!} {
\begin{tabular}{l l ccccccc} 
\toprule
Method & Language & Verdict & BLEU & EM & SM & DM & CB  & CA \\
\midrule
\multirow{2}{*}{Naive Copy} &Java & \xmark  & \textbf{80.28} & 0.0 & \textbf{84.22} & \textbf{53.64} & \textbf{75.43} & \textbf{89.93} \\ 
& Python & \xmark & \textbf{68.55} & 0.0  & \textbf{70.12} & \textbf{60.51} & \textbf{68.47} & \textbf{96.56} \\ 
\midrule
\multirow{4}{*}{PLBART} & Java & \xmark  & 58.49 & 0.45 & 66.92 & 43.08 & 57.23 & 31.36 \\
& Java & \checkmark  & 59.84 & 1.46 & 68.01 & 44.99 & 58.62 & 33.04 \\
& Python & \xmark  & 61.89 & 2.32 & 64.32 & 48.81 & 61.13 & 91.16 \\
& Python & \checkmark & 62.25 & 2.46 & 63.31 & 49.73 & 62.21 & 92.21 \\
\midrule
\multirow{4}{*}{CodeT5} & Java& \xmark  & 62.31 & \textbf{2.96} & 74.01 & 52.30 & 63.37 & 63.03 \\ 
& Java & \checkmark  & 62.54 & 2.45 & 73.93 & 53.29 & 63.71 & 64.23 \\ 
& Python & \xmark  & 64.92 & 2.74 & 68.79 & 56.21 & 63.53 & 92.80 \\ 
& Python & \checkmark & 64.67 & \textbf{2.97} & 68.45 & 56.04 & 63.28 & 92.70 \\
\bottomrule
\end{tabular}}
\caption{N-gram-based evaluation performances on \fixeval.
\textbf{EM} (Exact Match), \textbf{SM} (Syntax Match), \textbf{DM} (Dataflow Match), \textbf{CB} (CodeBLEU), and \textbf{CA} (Compilation Accuracy).
}
\label{tab:n_gram_results}
\end{table*}

 \begin{table*}[t!]
\centering
\resizebox{0.9\linewidth}{!} {%
\small
\begin{tabular}{l l c cccc  cccc}
    \toprule
    \multirow{2}{*}{Method} & \multirow{2}{*}{Language} & \multirow{2}{*}{Verdict} & \multicolumn{4}{c}{pass@$k$} & \multicolumn{4}{c}{top-$k$ TCA} \\
    \cmidrule{4-11}
    & & & $ k=1 $ & $ k=3$ & $k=5 $& $k=10$ & $k=1$ & $k=3$ & $k=5$ & $k=10$  \\
    \midrule
    \multirow{2}{*}{Naive Copy} & Java & - & 0.0 & - & - & - & 37.95 & - & - & - \\
    & Python & - & 0.0 & - & - & - & 41.55 & - & - & - \\

    \midrule
    \multirow{4}{*}{PLBART} & Java & \xmark & 7.51 & 14.21 & 17.32 & 24.14 & 39.89 & 33.02 & 31.88& 29.56\\ 
    &  Java & \checkmark & 8.43 & 17.65 & 21.51 & 27.17 & 43.87 & 37.78 & 34.71 & 32.78 \\ 
    
    & Python &\xmark & 6.15 & 12.59 & 15.74 & 19.98 & 49.81 & 40.79 & 37.63 & 34.43 \\ 
    & Python & \checkmark & 7.01 & 13.12 & 16.97 & 21.91 & 48.05 & 40.81 & 37.89 & 34.01 \\ 
    
    \midrule
    \multirow{4}{*}{CodeT5} & Java & \xmark & 8.65 & 15.62 & 19.63 & 24.44 & 41.00 & 34.00 & 32.70& 29.60\\ 
    &  Java & \checkmark & \textbf{10.94} & \textbf{18.77} & \textbf{22.66} & \textbf{27.96}& \textbf{44.99} & \textbf{38.80} & \textbf{35.87} & \textbf{32.90} \\ 
    & Python &\xmark & 6.86 & 13.07 & 16.27 & 20.51 & \textbf{50.20} & \textbf{41.20} & \textbf{38.50} & \textbf{35.20} \\ 
    & Python & \checkmark & \textbf{7.32} & \textbf{13.94} & \textbf{17.47} & \textbf{22.63} & 48.75 & 41.16 & 38.37 & 34.88 \\ 
    
    \bottomrule
\end{tabular}
}
\caption{Execution-based evaluation performance on \fixeval.}
\vspace{-2mm}
\label{tab:execution_based_results_1}
\end{table*}




\subsection{Evaluation Metrics}
To understand how accurately models perform on \textsc{FixEval}, we evaluate both conventional n-gram-based metrics and our proposed execution-based metrics, explained next.

\smallskip
\subsubsection{N-gram-based Metrics }
\label{n_gram_metrics}

\smallskip
\noindent\textbf{$\bullet$ Exact Match (EM)} indicates the percentage of the generated program fixes exactly match the reference programs.

\smallskip
\noindent\textbf{$\bullet$ BLEU} computes n-gram overlap between a model-generated fix and the reference. We use corpus-level BLEU scores~\cite{papineni2002bleu}. 

\smallskip
\noindent\textbf{$\bullet$ Syntax Match (SM)} represents the percentage of the sub-trees extracted from the Abstract Syntax Tree (AST) of a generated program that matches the subtrees in the AST of the reference programs. 

\smallskip
\noindent\textbf{$\bullet$ Dataflow Match (DM)} indicates the ratio of the number of matched candidate dataflows and the total number of reference dataflow~\cite{ren2020codebleu}.

\smallskip
\noindent\textbf{$\bullet$ CodeBLEU (CB)} is designed to measure the quality of a code with respect to a reference~\cite{ren2020codebleu}. Compared to BLEU, CodeBLEU also considers logical correctness based on an Abstract Syntax Tree (AST) in conjunction with data flow structure and grammatical similarity. 

\smallskip
\noindent\textbf{$\bullet$ Compilation Accuracy (CA)} indicates the percentage of generated program fixes that can be run without any compilation error. We use off-the-shelf compilers \texttt{javac} and \texttt{py\_compile}\footnote{\url{https://docs.python.org/3/library/py\_compile.html}} for Java and Python languages, respectively.

\smallskip
\subsubsection{Execution-based Metrics}
\smallskip
In program repair tasks, the input and output typically have high lexical overlapping. However, n-gram-based metrics may not accurately indicate the functional correctness of model-generated program fixes.
Further, a program can be fixed in multiple ways that differ from the reference program. Therefore, n-gram-based metrics may not be ideal for program repair evaluation.
Thus, we also evaluate \fixeval with execution-based metrics to alleviate these limitations.

Evaluating all generated programs on execution for all available test cases is memory-intensive and time-consuming. Therefore, we randomly select two program pairs per problem from the test split and evaluate them on all the available test cases. We select the data points with a similar distribution of the verdicts to ensure the evaluation data follows the actual distribution of the test data for different verdicts (AC, WA, TLE, etc.). Since our goal is not to exhaustively evaluate all models but to showcase the efficacy of the proposed dataset, we only evaluate CodeT5, the current state-of-the-art, on relevant tasks. We input the buggy code appended with the verdict information as input to the model and generate 10 outputs using beam search decoding. Then, we evaluate the output programs by running our test suite that simulates how online judges evaluate submitted programs. Our execution-based evaluation metrics, pass@$k$ and TCA@$k$, were introduced in recent works~\cite{kulal2019spoc,hendrycks2021measuring}. For self-containment, we provide the descriptions as follows:

\smallskip
\noindent\textbf{$\bullet$ Pass@k}~\cite{kulal2019spoc} evaluates functional correctness where $k$ code samples are generated per problem. A problem is considered solved if any sample passes all the unit tests, and the total fraction of problems solved is reported. However, this computation of pass@$k$ can have a high variance. Hence, we follow~\cite{chen2021evaluating} to compute pass@$k$, i.e., we generate $k \leq n$ samples per task (in this work, $n=10$ and $k \leq 10$), count the number of correct samples $c \leq n$ that pass all unit tests and calculate the unbiased estimator of pass@$k$ as follows.
\begin{equation}
    \text {pass@}k := \underset{ \mathcal{D}_{test}}{\mathbb{E}}\left[ 1 - \frac{{{n-c} \choose k}}{{n \choose k}} \right],
\end{equation}
where $\mathcal{D}_{test}$ denotes the \fixeval test set. Note that this is a strict measure, as a code repair is considered unsuccessful if a single failed test case exists. 

\smallskip
\noindent\textbf{$\bullet$ Test Case Average (TCA@k)}~\cite{hendrycks2021measuring} indicates the average number of test cases passed. Often, solutions successfully pass a subset of the test cases but fail to pass tests that cover corner cases. TCA@k allows for a less stringent model evaluation, as Pass@k accuracy may obscure model improvements.
Concretely, let $P$ be the set of problems in the test set and $|P|$ be the number of problems in $P$. Let the code fixes generated to solve problem $p \in P$ be denoted as $\left\langle code^{i}_p \right\rangle$, where $i$ denotes the index of generated fix and $k$ is the total number of generated fixes. Furthermore, let the set of test cases for problem $p$ be $\{(x_{p,c}, y_{p,c})\}_{c=1}^{|C_p|}$, where $x_{p,c}$ and $y_{p,c}$ are the input, output pair and $C_p$ is the number of available test case pairs for that problem. Then, the test case average for $k$ generated fixes (TCA@$k$) is
\begin{equation}
\scalemath{0.95}{
\frac{1}{|P|} \sum_{p \in P}
\frac{1}{k} \sum_{i=1}^{k}
\frac{1}{|C_{p}|} \sum_{c=1}^{|C_{p}|} 1\left\{\operatorname{eval}\left(\left\langle\operatorname{code}^{i}_{p}\right\rangle, x_{p, c}\right)=y_{p, c}\right\},
}
\end{equation}
where $\operatorname{eval}$ is the function evaluating a code fix in a test case by matching the output with the intended result. 

\subsection{Baselines} 
We consider the following two Transformer language models as the baseline methods:


\smallskip
\noindent\textbf{$\bullet$ PLBART}~\cite{ahmad-etal-2021-unified} is a BART~\cite{lewis2019bart} model trained on programming language corpora using three learning strategies: token masking, token deletion, and token infilling.

\smallskip
\noindent\textbf{$\bullet$ CodeT5} ~\cite{wang2021codet5} is a T5 model~\cite{raffel2019exploring} pretrained on programming languages via multiple objectives, such as span prediction and identifier tagging prediction. CodeT5 uses unimodal (code only) and bimodal (code text pairs) data for pretraining.


In addition, we consider \textbf{Naive Copy} as a baseline - the input buggy code is copied to the output (fixed code). 
Since typically there is a significant overlap between the buggy code and its fix, this baseline shows the minimum a model could achieve in n-gram-based evaluation metrics.

\smallskip
\noindent\textbf{Implementation:}
We finetune the base variants of PLBART and CodeT5 released by the respective authors on \fixeval programs. We finetune them with AdamW optimizer~\cite{loshchilov2019decoupled} with  $32$ batch size, $5 \times e^{-5}$ learning rate, early stopping with patience set to $3$ and $100$ warm-up steps. For inference, we use beam decoding with a beam size of $5$.

\section{Results}

We aim to address the following questions through our experiments and analyses:
\textbf{(1)} How well do pretrained Transformer models perform on \fixeval? and
\textbf{(2)} How do n-gram-based metrics track performance relative to execution-based evaluation? 
Our results validate the need for better program repair evaluation practices, demonstrating that \fixeval can fill a critical need in this line of research.




\subsection{N-gram-based Evaluation}
Table \ref{tab:n_gram_results} presents the n-gram-based evaluation results of all the baseline methods.
The Verdict column indicates the use of verdict information as conditional input when generating a candidate program fix. We observe that Naive Copy performs the best in all n-gram-based measures except for Exact Match (EM). This is because buggy and fixed code pairs typically overlap significantly.
Between the compared models, CodeT5 performs better than PLBART across all metrics and programming languages. We anticipate this is due to their identifier-aware pretraining that helps CodeT5 learn useful patterns.

Further, we observe a marginal performance increase in the baseline models with verdict information as conditional input for Java programs. However, there is no such indication for Python. We hypothesize that the lengthy description of the causes of test failures in Java positively impacts performance.
We further analyze the impact of verdicts in Section \ref{sec:verdict}.





\subsection{Execution-based Evaluation}
We compare n-gram-based and execution-based evaluations to check whether they correlate. 
The execution-based evaluation results are presented in Table~\ref{tab:execution_based_results_1}.
Naive Copy performed the best according to the n-gram-based metrics, and its performance in the execution-based evaluation demonstrates the limitations of n-gram-based evaluation. This suggests TCA and pass@k are better indicators for functional program correctness and evaluate models better than n-gram-based metrics.



\begin{figure}[t]
\centerline{\includegraphics[width=0.8\linewidth]{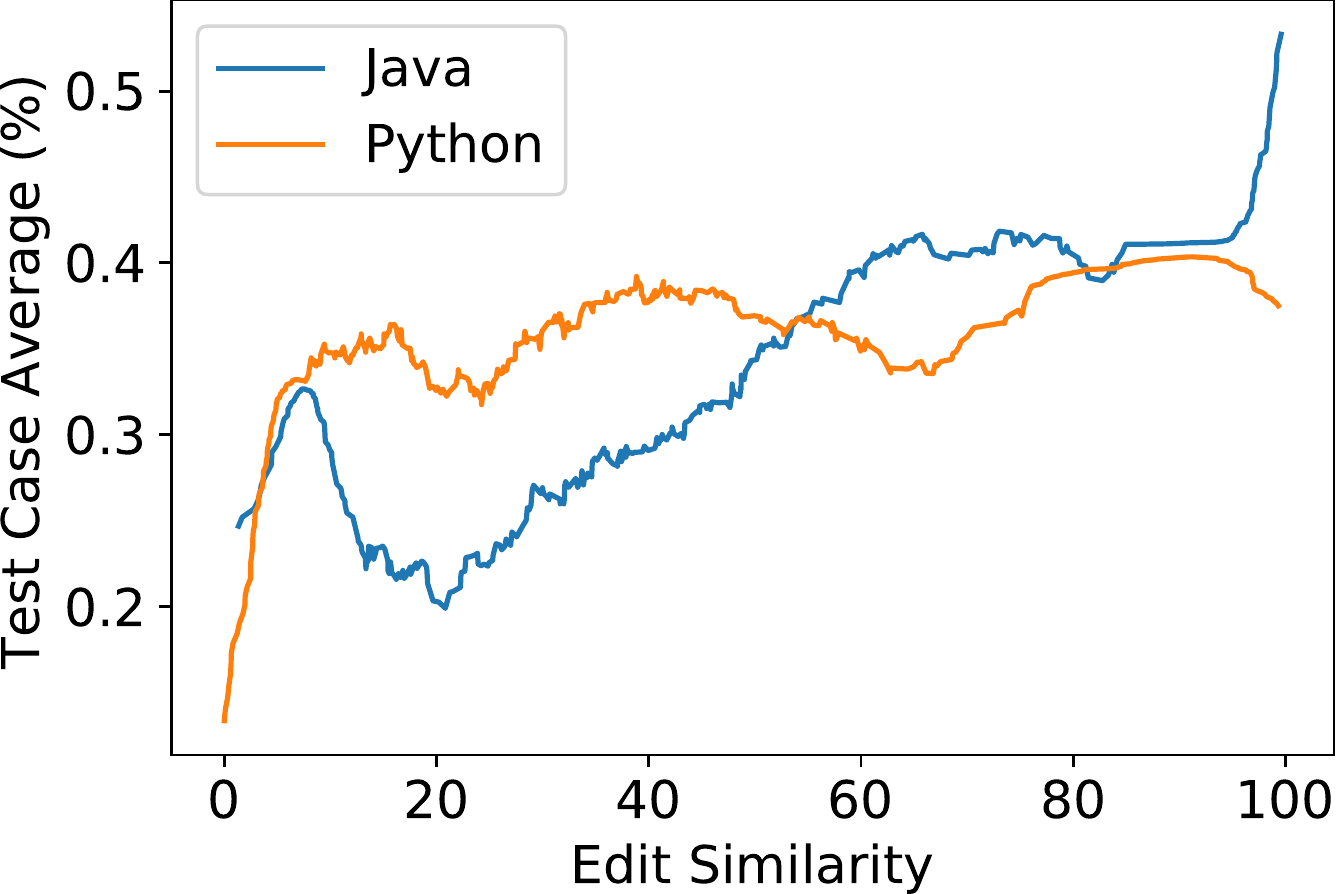}}
\caption{TCA increases as edit similarity between buggy and reference code increases for both Java and Python.}
\vspace{-2mm}
\label{fig:edit_sim_vs_tca}
\end{figure}

\section{Analysis}
We further analyze model performances on \fixeval to understand the effects of several components, e.g., verdict information, decoding algorithms, etc. All analyses are based on the CodeT5 model on sampled examples.

\subsection{Correlation to Edit Similarity} 
We analyze model performance based on edit similarity, assuming lower edit similarity between the buggy and fixed code indicates more difficulty in fixing the code. We sort all the data points on the \fixeval test set based on the buggy and reference (fixed) codes' edit similarity and plot the test case average of the CodeT5 model for both Java and Python.
In Figure \ref{fig:edit_sim_vs_tca}, we observe a mostly upward trending curve for Java, which indicates a positive correlation between edit similarity and model performance. However, no such correlation is evident for Python.

\begin{figure}[!t]
\centerline{\includegraphics[width=0.85\linewidth]{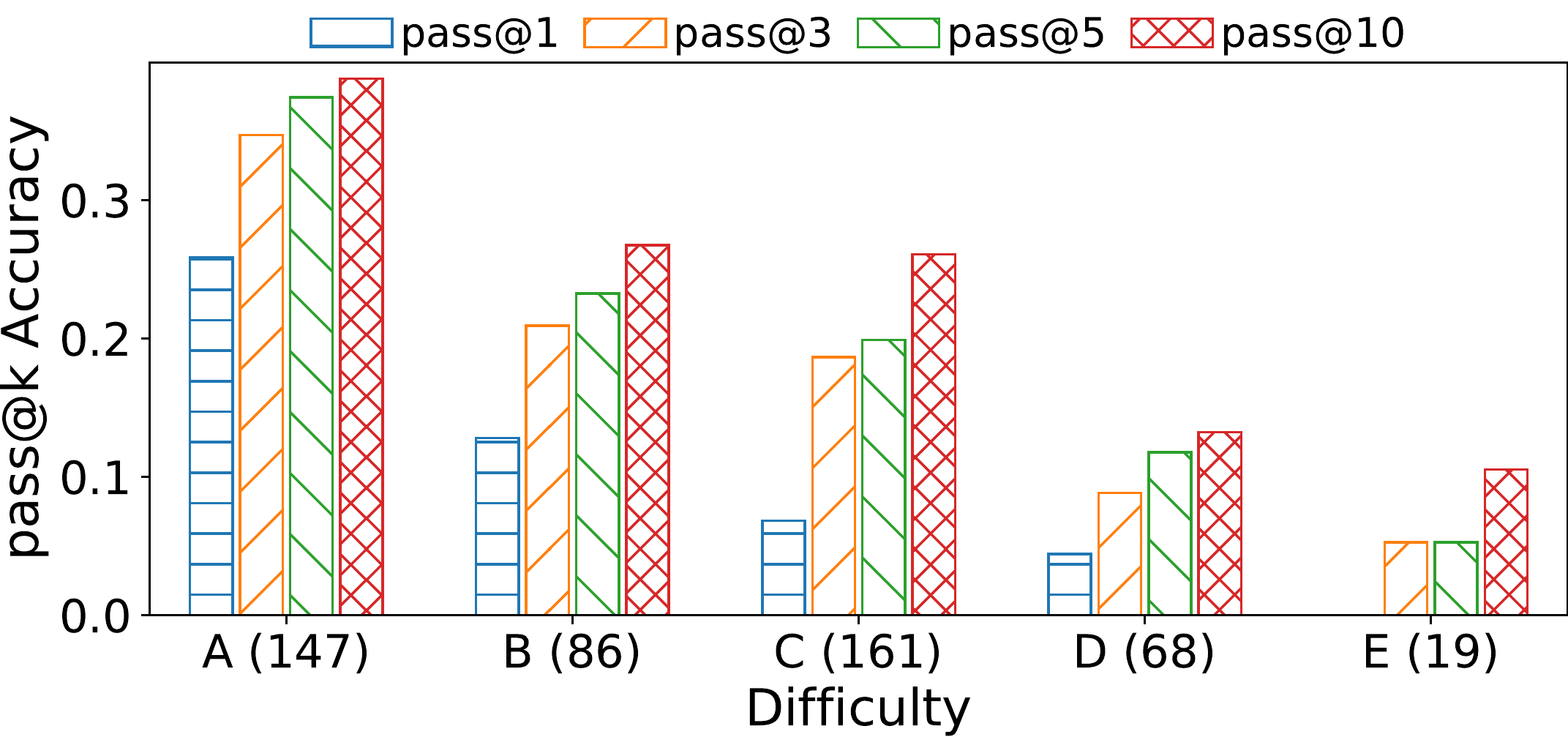}}
\caption{Pass@k accuracy (in Java) at different difficulty levels. Task labels A to E indicate increasing difficulty.}
\vspace{-2mm}
\label{fig:accept_by_difficulty}
\end{figure}

\begin{figure}[!t]
\centerline{\includegraphics[width=0.85\linewidth]{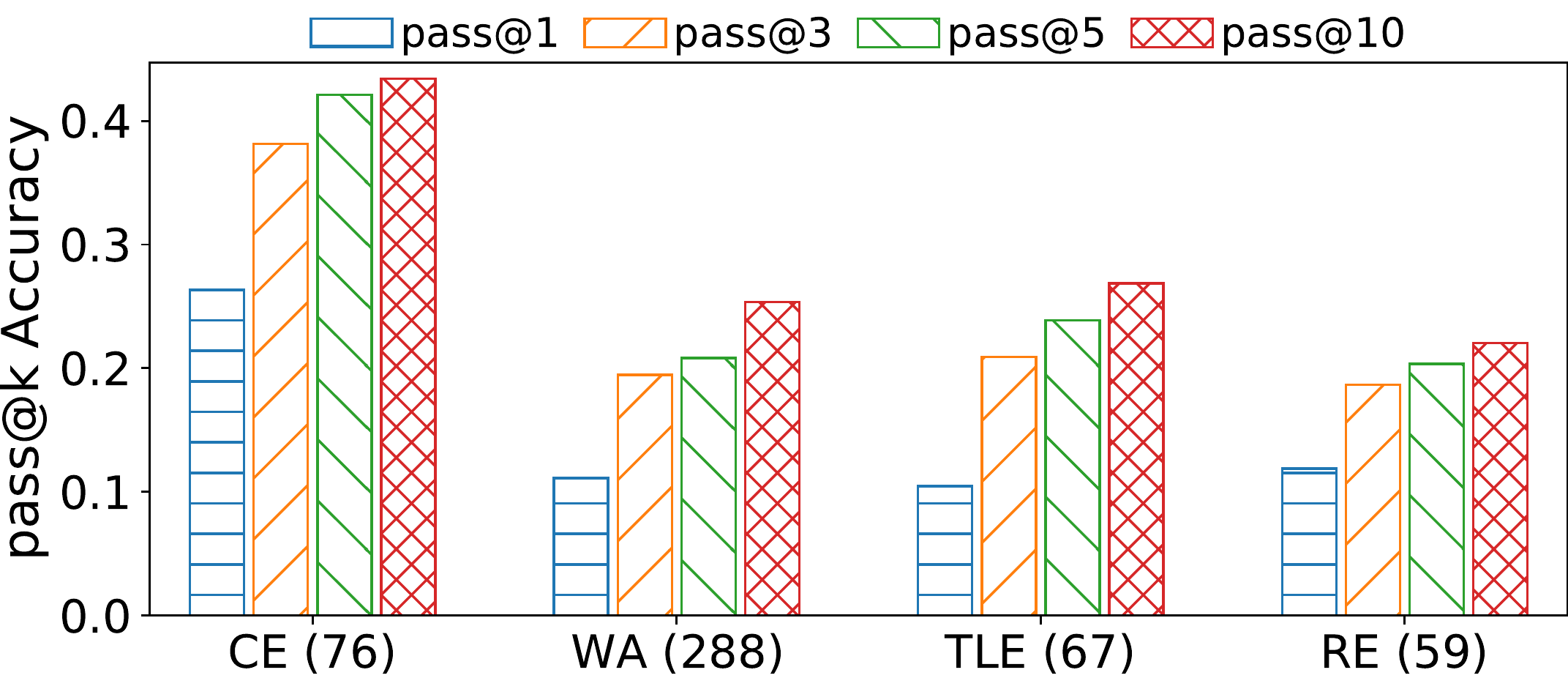}}
\caption{Pass@k accuracy (in Java) at different verdict labels, where CE = compilation error, WA = wrong answer, TLE = time limit exceeded, and RE = runtime error.}
\vspace{-2mm}
\label{fig:accept_by_verdict}
\end{figure}

\subsection{Correlation to Task Difficulty}
We analyze the effect of task difficulty as described in Section \ref{subsec:difficulty}.
From Figure \ref{fig:accept_by_difficulty}, we observe that the model performance degrades as difficulty increases (increasing difficulty from label A to E). 
This verifies our hypothesis that fixing a program that solves a challenging problem is harder.

\subsection{Correlation to Evaluation Verdict}
We analyze the effect of verdict type on performance. 
Figure \ref{fig:accept_by_verdict} shows that compilation errors (CE) are the easiest to solve as these mostly deal with syntactical changes to correct a program, whereas runtime errors (RE) or time limit exceeded errors (TLE) are much harder to fix since these indicate semantically incorrect code that requires multiple changes, sometimes the main logic of the algorithm.




\begin{figure}[t!]
    \centering
\includegraphics[width=0.8\linewidth]{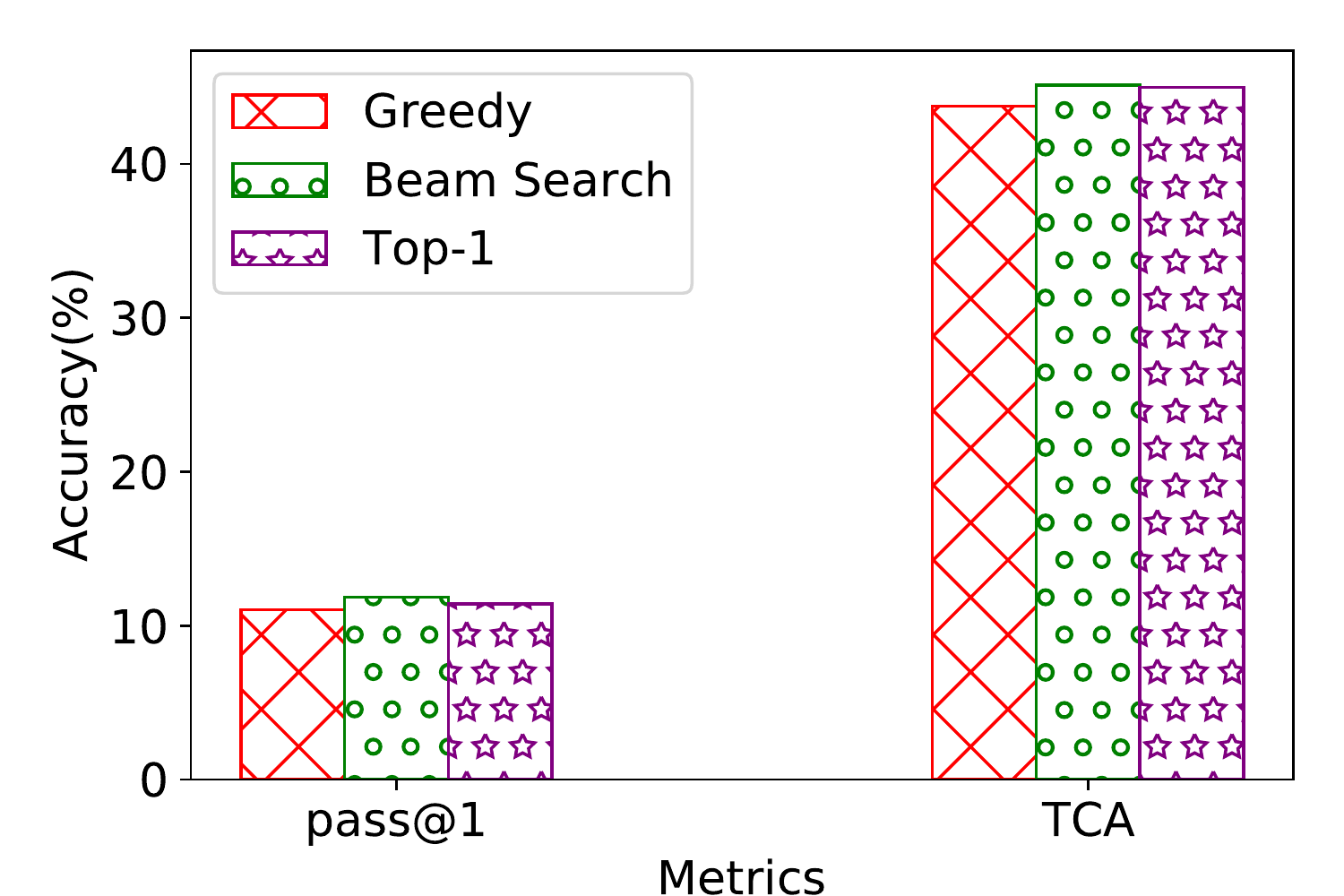}
\captionof{figure}{Pass@1 and TCA@1 accuracy for greedy decoding, beam search, and nucleus sampling. }
\label{fig:accept_by_Different_generation_types}
\vspace{-2mm}
\end{figure}

\subsection{Effect of Decoding Algorithms}
We use CodeT5 to generate fixed programs with various decoding strategies: (i) greedy, (ii) beam search with beam size $10$, and (iii) temperature-scaled nucleus sampling ($p=0.95$ and $t=0.7$). Figure  \ref{fig:accept_by_Different_generation_types} shows that beam search usually performs better than greedy decoding and sampling. We experiment with varying sampling temperatures from $0.2$ to $1.2$ and observe minor performance changes. We believe this is due to the nature of the problem, as the fixed program remains mostly similar to the buggy version, which results in the model remaining confident in its predictions. Hence, the temperature scaling doesn't result in substantial distribution changes.
\begin{figure*}[t!]
\centering

\vspace{5pt}
\begin{tabular}{l}
\begin{adjustbox}{valign=t,minipage=0.45\textwidth}
\begin{center}
    \underline{Buggy Program (verdict: Wrong Answer)}
\end{center}
\begin{tabular}{l}

\lstset{escapechar=~,style=CustomJava}
\begin{lstlisting}[ 
    linebackgroundcolor={%
    \ifnum\value{lstnumber}=6
        \color{red!10}
    \fi
    \ifnum\value{lstnumber}=7
        \color{red!10}
    \fi
    }
]
import java.util.*;
import java.lang.*;
public class Main {
    public static void main(String[] args){
        Scanner sc = new Scanner(System.in);
        int a = sc.nextInt();
        int b = sc.nextInt();
        long ans = a*b/gcd(a, b);
        System.out.println(ans);
        sc.close();
    }
    public static long gcd(long m,long n){
        if (m < n) return gcd(n, m);
        if (n==0) return m;
        return gcd(n, m % n);
    }
}
\end{lstlisting}
\end{tabular}
\end{adjustbox}
\hspace{8pt}
\begin{adjustbox}{valign=t,minipage=0.45\textwidth}
\begin{center}
    \underline{Fixed Program}
\end{center} 
\begin{tabular}{l}
\lstset{escapechar=@,style=CustomJava}
\begin{lstlisting}[ 
    linebackgroundcolor={%
    \ifnum\value{lstnumber}=6
            \color{green!10}
    \fi
    \ifnum\value{lstnumber}=7
            \color{green!10}
    \fi
    }
]
import java.util.*;
import java.lang.*;
public class Main{
    public static void main(String[] args){
        Scanner sc = new Scanner(System.in);
        long a = sc.nextInt();
        long b = sc.nextInt();
        long ans = a*b/gcd(a, b);
        System.out.println(ans);
        sc.close();
    }
    public static long gcd(long m,long n){
        if (m < n) return gcd(n, m);
        if (n==0) return m;
        return gcd(n, m % n);
    }
}
\end{lstlisting}

\end{tabular}
\end{adjustbox}
\end{tabular}

\centering

\vspace{5pt}
\begin{tabular}{l}
\begin{adjustbox}{valign=t,minipage=0.45\textwidth}
\begin{center}
    \underline{Buggy Program (verdict: Compilation Error)}
\end{center}
\begin{tabular}{l}

\lstset{escapechar=~,style=CustomJava}
\begin{lstlisting}[ 
    linebackgroundcolor={%
    \ifnum\value{lstnumber}=7
        \color{red!10}
    \fi
    \ifnum\value{lstnumber}=8
        \color{red!10}
    \fi
    }
]
import java.util.*;
public class Main {
    public static void main(String[] args){
        Scanner sc = new Scanner(System.in);
        int a = sc.nextInt();
        int b = sc.nextInt();
        if((A - B) % 2 == 0){
            System.out.println((A + B)/2);
        }
        else {
            System.out.println("IMPOSSIBLE");
        }
    }
}
\end{lstlisting}
\end{tabular}
\end{adjustbox}
\hspace{8pt}
\begin{adjustbox}{valign=t,minipage=0.45\textwidth}
\begin{center}
    \underline{Fixed Program}
\end{center} 
\begin{tabular}{l}
\lstset{escapechar=@,style=CustomJava}
\begin{lstlisting}[ 
    linebackgroundcolor={%
    \ifnum\value{lstnumber}=7
            \color{green!10}
    \fi
    \ifnum\value{lstnumber}=8
            \color{green!10}
    \fi
    }
]
import java.util.*;
public class Main {
    public static void main(String[] args){
        Scanner sc = new Scanner(System.in);
        int a = sc.nextInt();
        int b = sc.nextInt();
        if((a-b) % 2 == 0) {
            System.out.println((a+b)/2);
        }
        else {
            System.out.println("IMPOSSIBLE");
        }
    }
}


\end{lstlisting}

\end{tabular}
\end{adjustbox}
\end{tabular}

\caption{
Examples of successful fixes of buggy programs in Java when the verdict information is provided as additional model input. 
Buggy and fixed statements are marked in red and green, respectively.
}
\vspace{-2mm}
\label{fig:example_2}
\end{figure*}

\subsection{Effect of Modeling Verdict}
\label{sec:verdict}
We analyze the test examples repaired correctly when the model has access to the verdict information. We observe that verdict information helps in fixing some instances of buggy code. We further observe that the model without verdict information attempts to add unnecessary but syntactically correct code snippets that cannot fix the actual error. In contrast, the model with verdict information can pinpoint the exact location of the error and make the code more consistent. We provide relevant examples in Figure \ref{fig:example_2}. 

\subsection{Summary of Analysis}
We study performance trends concerning edit similarity, task difficulty, and evaluation verdicts to show that some bug-fixing tasks are trivial while many are challenging. Therefore, we encourage future work to consider all aforementioned aspects while performing evaluations. The choice of decoding strategy produces marginal differences; therefore, it is not a crucial factor in improving bug-fixing models. We encourage future works to study feedback-based (e.g., feedback from an oracle) approaches to improve bug-fixing models.

\subsection{Limitations}
We acknowledge that fixing programming problem solutions does reflect real-world software bugs from professional developers. Therefore, more effort is necessary to develop benchmarks that simulate bug fixing of real-world software programs to evaluate automated program repair models.







\section{Conclusion}
We introduce \fixeval, a dataset of program fixes to facilitate bug-fixing model development and evaluation. 
We benchmark two pretrained language models on \fixeval and showcase that traditional evaluation metrics are sub-optimal compared to execution-based metrics that capture contextual program repair requirements often found in practice. 

We believe \fixeval could help facilitate several software engineering tasks, such as code completion, code editing, code search, verdict-conditioned code repair, verdict prediction, and chain edit suggestion tasks. In the future, since the provided test cases are language-independent, our work can be extended to other programming languages.

\bibliographystyle{abbrv}
\bibliography{biblio.bib}


\end{document}